\documentclass[twocolumn,tighten,A4paper]{aastex61}

\begin{document}

\def\degr{\ensuremath{^\circ}}
\def\arcmin{\ensuremath{^\prime}}
\def\arcsec{\ensuremath{^{\prime\prime}}}
\def\kms{\rm km\,s^{-1}}



\title{Radio interferometric observation of an asteroid occultation}


\author[0000-0002-1189-9790]{Jorma Harju}
\affiliation{Department of Physics, P.O. BOX 64, FI-00014 University
  of Helsinki, Finland}
\affiliation{Max-Planck-Institut f\"ur extraterrestrische Physik,
  Gie{\ss}enbachstra{\ss}e 1, D-85748 Garching, Germany}

\author[0000-0002-1004-8898]{Kimmo Lehtinen}
\affiliation{Finnish Geospatial Research Institute FGI, 
Geodeetinrinne 2, FI-02430 Masala, Finland}

\author[0000-0003-4410-5421]{Jonathan Romney} 
\affiliation{Long Baseline Observatory, 1003 Lopezville Road, Socorro,
  NM 87801, U.S.A}

\author{Leonid Petrov}
\affiliation{Astrogeo Center, Falls Church, U.S.A.}

\author[0000-0002-5624-1888]{Mikael Granvik}
\affiliation{Department of Computer Science, Electrical and Space
  Engineering, Lule\aa{} University of Technology, Box 848, SE-98128
  Kiruna, Sweden}
\affiliation{Department of Physics, P.O. BOX 64, FI-00014 University
  of Helsinki, Finland}

\author{Karri Muinonen}
\affiliation{Department of Physics, P.O. BOX 64, FI-00014 University
  of Helsinki, Finland}
\affiliation{Finnish Geospatial Research Institute FGI, 
Geodeetinrinne 2, FI-02430 Masala, Finland}

\author[0000-0002-7722-8412]{Uwe Bach}
\affiliation{Max-Planck-Institut f\"ur Radioastronomie,
  Auf dem H\"ugel 69, D-53121 Bonn, Germany}

\author{Markku Poutanen}
\affiliation{Finnish Geospatial Research Institute FGI, 
Geodeetinrinne 2, FI-02430 Masala, Finland}

\begin{abstract}

  The occultation of the radio galaxy 0141+268 by the asteroid (372)
  Palma on 2017 May 15 was observed using six antennas of the Very
  Long Baseline Array (VLBA).  The shadow of Palma crossed the
    VLBA station at Brewster, Washington. Owing to the
  wavelength used, and the size and the distance of the asteroid, a
  diffraction pattern in the Fraunhofer regime was observed. The
  measurement retrieves both the amplitude and the phase
  of the diffracted electromagnetic wave. This is the first
  astronomical measurement of the phase shift caused by
  diffraction. The maximum phase shift is sensitive to the effective
  diameter of the asteroid. The bright spot at the shadow's center,
  the so called Arago--Poisson spot, is clearly detected in the
  amplitude time-series, and its strength is a good indicator of the
  closest angular distance between the center of the asteroid and the
  radio source. A sample of random shapes constructed using a
  Markov chain Monte Carlo algorithm suggests that the silhouette of
  Palma deviates from a perfect circle by $26\pm13\%$.  The
    best-fitting random shapes resemble each other, and we suggest
    their average approximates the shape of the silhouette at the
    time of the occultation. The effective diameter obtained for
  Palma, $192.1\pm 4.8$\,km, is in excellent agreement with recent
  estimates from thermal modeling of mid-infrared photometry. Finally,
  our computations show that because of the high positional accuracy,
  a single radio interferometric occultation measurement can reduce
  the long-term ephemeris uncertainty by an order of magnitude.
  
\end{abstract}


\nopagebreak

\section{Introduction} \label{sec:intro}

Observations of lunar occultations provided the first accurate
positions for compact extragalactic radio sources, and contributed to
the discovery of quasars (\citealt{1963Natur.197.1037H};
\citealt{1963Natur.197.1040O}; \citealt{1963Natur.197.1040S}).  Lunar
occultations have also been used to derive high-resolution images of
radio sources from the Fresnel diffraction fringes observed with
single-dish telescopes (e.g., \citealt{1967ApJ...148..669H};
\citealt{1979ApJ...234..485M}; \citealt{1980ApJ...235L..33S};
\citealt{1994ApJ...423L.143C}).  Restoring techniques developed by
\cite{1962AuJPh..15..333S} and \cite{1964ApJ...140...65V} were
utilized in these works. More recently, the brightness distributions
of compact radio sources have mainly been studied using
interferometric arrays such as Very Long Baseline Interferometry
(VLBI). Nevertherless, radio occultations by solar system bodies
remain useful for determining the properties of the foreground
objects, including planetary atmospheres and coronal mass ejections
from the Sun (\citealt{2017SoPh..292...56K};
\citealt{2017ApJ...836..114W}).  The basic principles and methods of
radio (lunar) occultation measurements were described by
\cite{1976MOEP..12C...92H}.

One application of radio occultations is the possibility of
determining the size of an asteroid by a single, short
measurement. This measurement also gives an accurate position of
  the asteroid at the time of the occultation, and can be used to
  constrain its shape, particularly when the angular size of the object is too
  small to be imaged with any currently available instruments.  The
  radio occultation method for asteroids was first demonstrated by
  \cite{2016ApJ...822L..21L} who used the 100m Effelsberg telescope
  to observe the occultation of a radio galaxy by the asteroid (115)
  Thyra.  As discussed by \cite{2016ApJ...822L..21L}, at radio
  wavelengths the shadow of the asteroid is often dominated by
  diffraction fringes. The observed diffraction pattern depends on the
  wavelength, and the size and the distance of the asteroid. A rule of
  thumb is that a sharp shadow is observed when the Fresnel number,
  $F$, is greater than 1, whereas a diffraction pattern in the
  Fraunhofer regime is observed for $F\leq 1$ (e.g.,
  \citealt{2014ApOpt..53.3540T}). In the latter case the diffraction
  pattern is much larger than the geometrical shadow. The Fresnel
  number is defined by $F = a^2/(d \lambda)$, where $a$ and $d$ are
  the radius and the distance of the occluder, and $\lambda$ is the
  wavelength.

  The astrophysical interest in asteroid size determination lies,
    for example, in the fact that it is needed for estimating the bulk
    density, which in turn contains information of the composition and
    internal structure of an asteroid \citep{2012P&SS...73...98C}. So
    far, only a third of the density estimates are more precise than
    20\% \citep{2012P&SS...73...98C}.  Increasing the sample of
    asteroids with accurate density estimates is important given the
    large variation in asteroid diameters (and hence in self-gravity),
    as well as the potentially large variation in interior structure,
    which reflects the collisional history of the solar system. Another
    challenge is the determination of the mass, which for an isolated
    asteroid, is based on orbital deflections during close encounters
    with other asteroids.  In the fortuitous situation in which an 
    asteroid has a satellite, its orbit around the primary body can be
    used to substantially constrain the mass.  Radio occultation
    observations can help to discover binary systems by virtue of
    diffraction fringes from the satellite.

    Here, we report on an asteroid occultation observation using the
    Very Long Baseline Array (VLBA; \citealt{1994IEEEP..82..658N}),
    operated by the Long Baseline Observatory. During an occultation,
    the direct path to the source blocked and diffracted
    electromagnetic radiation arrives at the receiver. The ampltitude
    and phase of the diffracted wave are different from those measured
    for the unobstructed wave.  In contrast to most optical
    interferometry, in which the correlated signal power is detected
    (for example, in a CCD), radio interferometers correlate the
    voltages from pairs of telescopes, and are able to measure the
    phase as well as amplitude of the correlated signals. In the
    course of calibration and imaging, one can correct for both phase
    and amplitude variations and recover the complex effects of the
    occultation.  This offers important advantages over single-dish
    measurements, as two quantities characterizing the diffracted
    wavefront are measured simultaneously, and, as shown in this
    paper, the phase is more sensitive to the asteroid size than the
    amplitude. The occultation observation of
    \cite{2016ApJ...822L..21L} with a single total power receiver had
    problems with reconciling the expected diffraction pattern from a
    contiguous occluder of the size of Thyra. In particular, the
    brightness of the first diffraction maxima could not be
    explained. In the present work, the measured amplitude and phase
    curves can be satisfactorily explained in terms of the
    Fresnel-Kirchoff diffraction theory, and lead to a plausible model
    for the asteroid silhouette.

\begin{figure}[htbp]
  \figurenum{1} 
\unitlength=1mm
\begin{picture}(80,65)
\put(0,0){
\begin{picture}(0,0) 
\includegraphics[width=8.0cm]{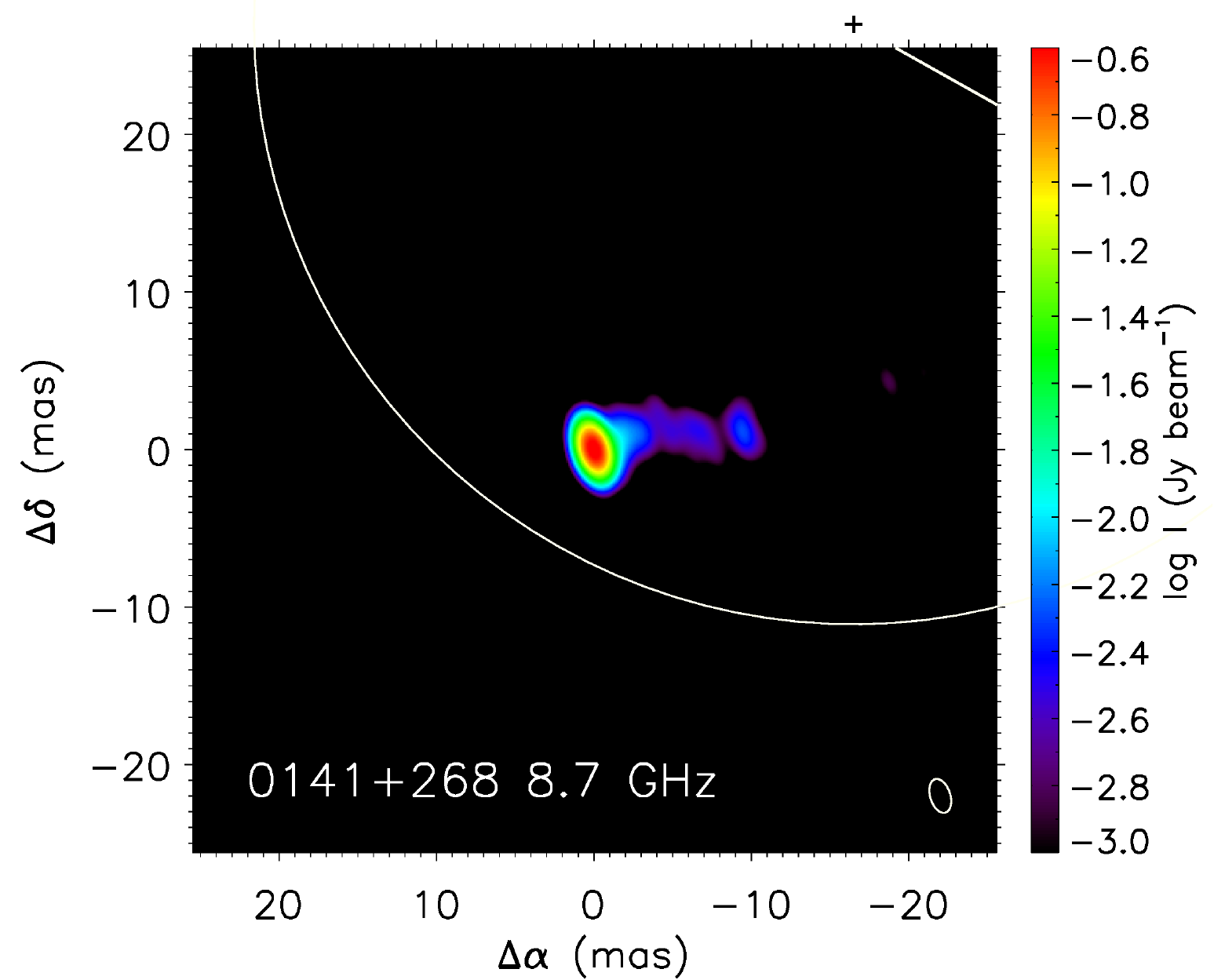}
\end{picture}}
\end{picture}
\caption{Image of the active galactic nucleus 0141+268 (J0144+270) at
  8.7 GHz from the VLBA survey VCS-II \citep{2016AJ....151..154G}. The
  synthesized beam size of the observation is shown in the bottom
  right ($2.2\times1.3$ mas, P.A. $17\fdg4$). The predicted closest
  position of (372) Palma as seen from Brewster is indicated with a
  plus sign (outside the frame). The white arc marks the approximate
  silhouette of the asteroid, assuming that it is a circular disk with
  a diameter of 190\,km (76 mas).}
\label{fig:0141+268}
\end{figure}

The background source in the present observations is the
active galactic nucleus (AGN) 0141+268 (J0144+270), associated with a
BL Lac type galaxy \citep{2016AJ....152...12L}. At the wavelength used
($\lambda=4.2$ cm), the source consists of a compact nucleus, $< 3$
milliarcseconds (mas), and a faint jet. An image of the source at 8.7
GHz from the VLBA experiment VCS-II-D/BG219D (2014 June 9;
\citealt{2016AJ....151..154G}), is shown in
Figure~\ref{fig:0141+268}.\footnote{Other VLBA images and the radio
  spectrum of the source can be found at \url{www.physics.purdue.edu/MOJAVE}
and at \url{astrogeo.org/vlbi_images}}.  

  The occluder is the asteroid (372) Palma, which resides in the outer
  parts of the main asteroid belt (semimajor axis $a=3.15$\,au,
  eccentricity $e=0.26$, inclination $i=23\fdg8$). The SMASSII
  spectroscopic classification of Palma is B
  \citep{2002Icar..158..146B}, geometric albedo $p_{\rm
    V}=0.059\pm0.009$ \citep{2012ApJ...759L...8M}, and bulk density
  $1.40\pm0.18$\,g\,cm$^{-3}$ \citep{2012P&SS...73...98C}. Palma is
  not associated with any of the currently known asteroid families.
Recent estimates for its diameter from thermal modeling of
mid-infrared photometric data range from 187 to 191\,km
(\citealt{2012P&SS...73...98C}; \citealt{2014ApJ...791..121M}).

The paper is organized as follows. In Section 2, we present the
prediction for the occultation observable with one of the VLBA antennas, 
located in Brewster, Washington, USA. In Section 3 we 
describe the VLBA observations. In Section 4, the observed visibilities
are interpreted in terms of the Fresnel-Kirchhoff diffraction theory
applied to different silhouette models. The results are discussed in
Section 5.

\begin{figure}[htbp]
  \figurenum{2} 
\unitlength=1mm
\begin{picture}(80,75)
\put(0,0){
\begin{picture}(0,0) 
\includegraphics[width=8.0cm]{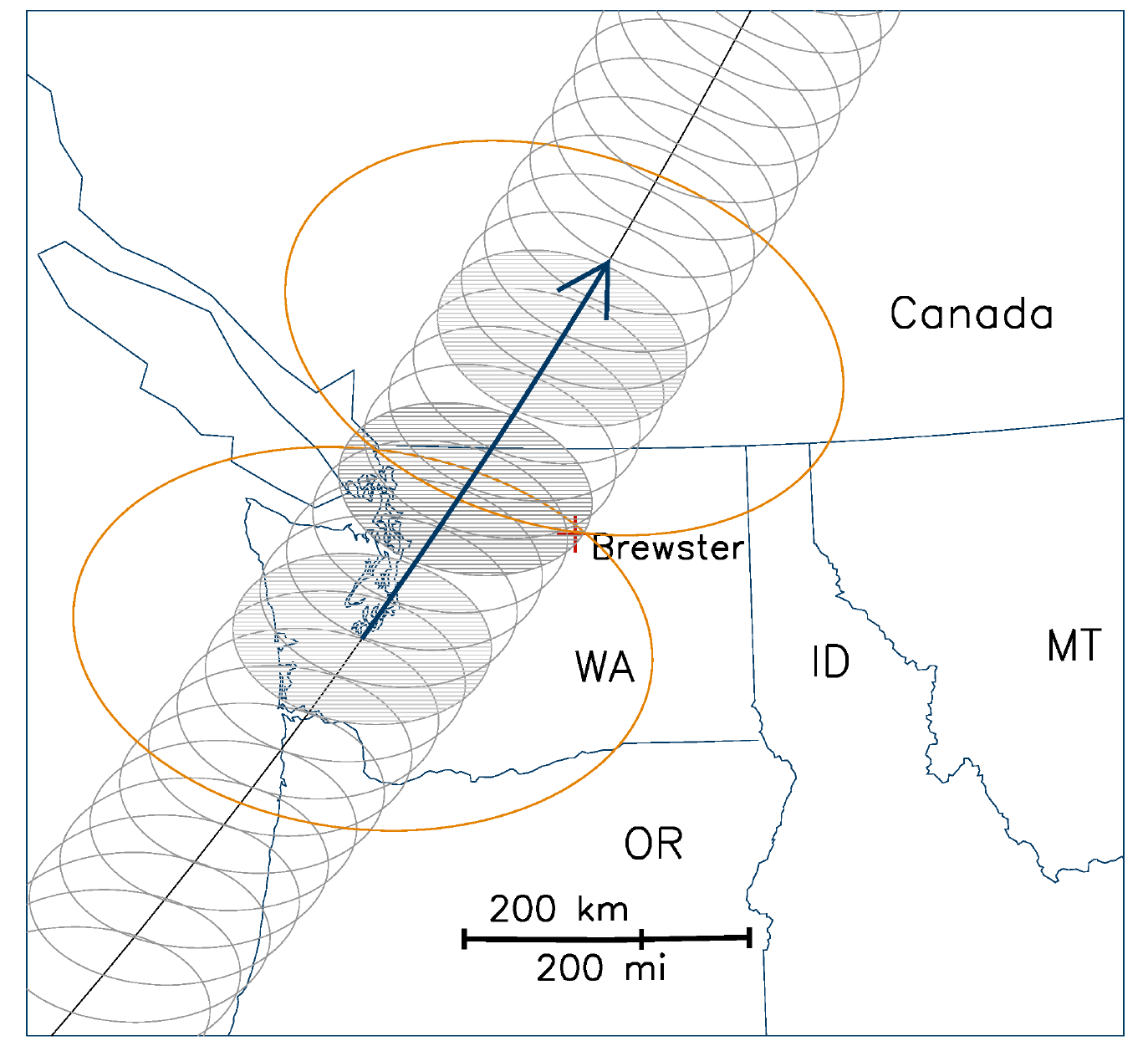}
\end{picture}}
\end{picture}
  \caption{Shadow path of (372) Palma on 2017 May 15, near Brewster,
    Washington. The arrow shows the distance the
    shadow traveled in 10 seconds (510\,km). The projections of the
    geometric shadow of the asteroid are shown with small ellipses at
    intervals of 1 second. The asteroid silhouette is approximated
    by a circular disk with a radius of 96\,km. The two bigger ellipses
    show the first maxima of the diffracted intensity pattern outside
    the geometric shadow, $4$~s before and after its closest approach
    to Brewster. The prediction for the size of the bright rings 
    is based on the 96\,km circular disk.}
\label{fig:palma_shadow}
\end{figure}

\section{Occultation Prediction}

Palma was predicted to occult the AGN 0141+268 (J0144+270) on 2017 May 15,
as seen from the Brewster VLBA station ({\sc br-vlba}). 
The predictions were made using the LinOccult
program\footnote{\url{http://andyplekhanov.narod.ru/occult/occult.htm}}.
LinOccult requires three auxiliary data sets. (1) The position of the
background object was adopted from the latest Radio Fundamental
Catalog (RFC), available at \url{http://astrogeo.org/rfc}
(\citealt{2008AJ....136..580P}; \citealt{2017MNRAS.467L..71P}). (2)
The orbital elements of the asteroid were taken from a database
provided by Lowell
Observatory\footnote{\url{ftp://ftp.lowell.edu/pub/elgb/astorb.html}}. (3)
The ephemerides of the major planets were obtained using the JPL
HORIZONS
service\footnote{\url{ftp://ssd.jpl.nasa.gov/pub/eph/planets/Linux/de405},
  file lnxp1600p2200.405}. 　

The asteroid radius ($a\sim 96$\,km;
\citealt{2012P&SS...73...98C}; \citealt{2014ApJ...791..121M}), and its
distance at the time of the observation ($d\sim 3.436$ au) imply a
Fresnel number of 0.43 at $\lambda=4.2$ cm. This means that a
Fraunhofer diffraction pattern was observable on the Earth. The
predicted shadow path in the vicinity of Brewster is shown in
Figure~\ref{fig:palma_shadow}. According to this prediction, the
ground speed of the shadow was 51\,km\,s$^{-1}$ near Brewster, and the
telescope was located just inside the geometric shadow at the time of
the closest approach, UT 14:31:23. At this time, the asteroid and the
radio galaxy were in the east, at an elevation of $42\degr$. The
projection of the shadow on the Earth is therefore elongated in an
east-west direction. Also shown in this figure are the projections of
the first bright ring of the diffracted intensity pattern, 4\,s 
before and after the deepest occultation. The radius of the
  bright ring (in a plane perpendicular to the line-of-sight to the
  source) is $\sim 215$\,km, calculated assuming a circular occluder
  with a radius of 96\,km. The detectable diffraction fringes are
  expected to extend to a distance approximately twice this
  radius. Taking the projection effect into account, the maximum
  distance from the shadow center where the diffraction fringes could
  possibly be detected is $\sim 630$\,km. This means that other VLBA
antennas than Brewster received unobscured signal from 0141+268.

\section{Very Long Baseline Array Observations}

\newcommand{\Frac}[2]{\frac{\displaystyle\strut #1}{\displaystyle\strut #2} }

The occultation was observed by six VLBA antennas: {\sc br-vlba}
(Brewster), {\sc fd-vlba} (Fort Davis), {\sc kp-vlba} (Kitt Peak),
{\sc la-vlba} (Los Alamos), {\sc pt-vlba} (Pie Town), and {\sc
  ov-vlba} (Owens Valley) located in the continental United States.
The lengths of baselines to {\sc br-vlba} ranged from 1.2 to 2.3
thousand kilometers. The central frequency of the observations was
6.996 GHz, and the total bandwidth was 256 MHz. Both right and left
circular polarizations were recorded with 2 bits per sample.

The observations started at UT 14:00, on May 15, with a five minute scan
on the strong radio source 0149+218, which was used as the fringe finder and
bandpass calibrator. Thereafter, the antennas tracked 0141+268 until UT 15:00
with 10~s long pauses every 5 minutes. The occultation was predicted to
be deepest at 14:31:23. System temperature ranged from 25 to 35~K, and
the system equivalent flux density was in the range 240--290~Jy.

The data were correlated in the Science Operations Center in Socorro,
using NRAO's implementation of the DiFX (Distributed FX) software
correlator (\citealt{2011PASP..123..275D}; FX means that the Fourier
transform is applied before the cross-multiplication of signals).  The
correlator integration time was 0.25~s and the frequency resolution
was 250~kHz.  Further processing was done with the ${\cal PIMA}$ VLBI
data analysis software package\footnote{Documentation is available at
  \url{http://astrogeo.org/pima}}.  The data were split into 290\,s
long segments, and the residual phase delay and group delays were fit
into the spectrum of the cross-correlation function produced by the
correlator, also known as the fringe visibility. The fringe
  visibilities were counter-rotated to the contribution of the group
  delay and the phase delay rate, and coherently averaged over
  frequency. For details on the fringe fitting procedure, we refer the
  reader to \cite{2011AJ....142...35P} and \cite{Thompson2017}.

Only one antenna, {\sc br-vlba} was affected by the
  occultation. During the occultation, the power of the signal
  recorded at {\sc br-vlba} dropped, and the optical path delay from
  the source to the antenna changed with respect to the unobscured
  situation. Even though the $(u,v)$ visibility coordinates for the
  five baselines are different, and one can expect differences in the
  amplitudes and, especially, the phases between these, the changes in the
  normalized amplitudes and the relative phases during the occultation
  should be similar for all of them. For the purposes of the present
  study we only need to retrieve these relative changes. Therefore,
  after the fringe fitting, we computed arithmetic averages over the
  five baselines. The residual fringe phases were stacked with zero
  mean, and the stacked fringe amplitudes were normalized to unity.
  The change in the path delay at {\sc br-vlba} propagates directly to
  the visibility phases, $\Phi$, for all the five baselines including
  Brewster.  In contrast, the change in the normalized visibility
  amplitude, $|V|$, during the occultation is proportional to the
  square root of the relative power drop measured at {\sc br-vlba},
  that is, $|V| \propto \sqrt{\frac{\Delta P_{\rm B}}{P_{\rm B}}}$.
  The effects removed in the calibration process occur on long
  timescales compared with the brief occultation event.

\begin{figure}[htbp]
\figurenum{3}
\unitlength=1mm
\begin{picture}(80,85)
\put(0,0){
\begin{picture}(0,0) 
\includegraphics[width=8.0cm]{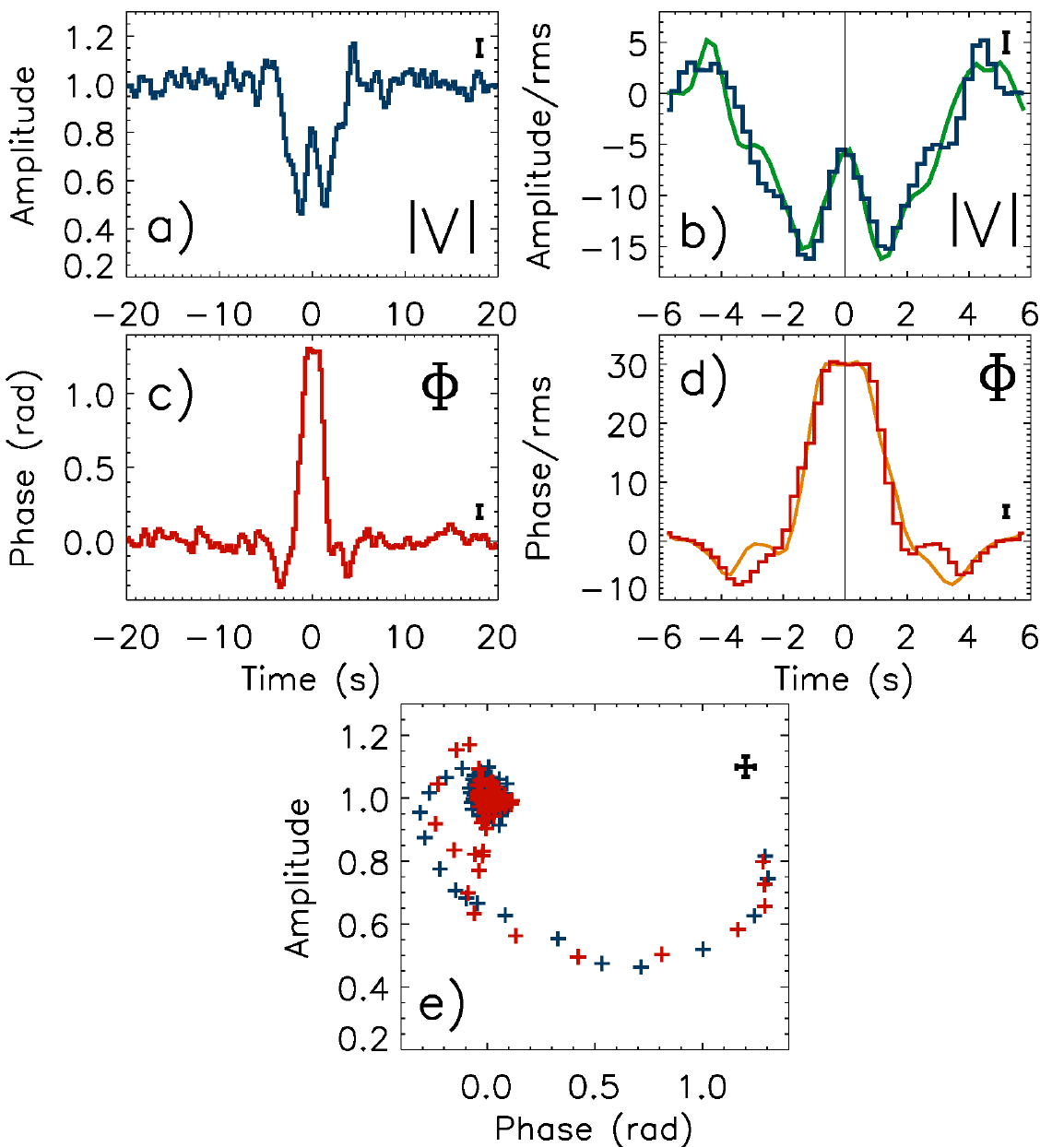}
\end{picture}}
\end{picture}
\caption{Stacked visibility amplitudes, $<|V|>$ (a), and phases,
  $<\Phi>$ (c), during the occultation. The vertical bars show
    the rms scatter of the normalized amplitude (a) and the phase (c)
    (calculated excluding a 20~s period around the occultation). The
  averaging is performed over five VLBA baselines including Brewster.
  The time-series are smoothed to a resolution of 0.35~s.  The
  amplitude of the unobscured signal is normalized to unity. The right
  panels ((b) and (d)) show 12~s portions of the amplitude and phase
  time-series in terms of the rms scatter (vertical
    bars). Time-inverted amplitude and phase curves (green and orange
  curves) are superposed to highlight asymmetries.  The bottom panel
  (e) shows the amplitude vs. phase diagram. The points before and
  after the deepest occultation are indicated with blue and red,
  respectively. The cross shows the rms scatters of the two
    quantities.}
\label{fig:aver_ampls+phases}
\end{figure}

For individual baselines, the rms noise levels over the 290~s
  interval, including the occultation (excluding 20~s around the
  closest approach), range from 0.112 to 0.147 for the normalized
  amplitudes, and from 0.128 to 160 rad for the phases.  The rms
  values of the stacked amplitudes and phases for the same period are
  0.079 and 0.083 rad, respectively.  In order to reduce the scatter
  in residual fringe amplitudes and phases, at the expense of the time
  resolution, we applied a weak Gaussian filter using the kernel
  $K(t,t_0)= \exp\left\{-(t-t_0)^2/(2a^2)\right\}$, where $t$ is time
  and the parameter has the value $a = 0.35$~s.

The stacked fringe phases exhibit fluctuations of $\sim\! 0.1$
  rad on time scales of 50--100\,s due to changes in the atmosphere
  path delay and possible phase and frequency offsets of the station
  clocks. In order to alleviate the contribution of these smooth
  fluctuations, we first fitted the phase pattern during the
  occultation by the {\sl sinc} function, and subtracted this model from
  the time-series. Therafter, a low-pass Gaussian filter with $a =
  10$~s was applied to the residuals. Finally, the smoothed, 
low-pass-filtered phases were subtracted from those filtered with $a = 0.35$~s.
  These residual phases and amplitudes are used in the subsequent
  analysis.  The rms scatter of this 270~s long dataset, with $\pm
  10$~s around the occultation excluded, is 0.033 for the normalized
  amplitude and 0.043~rad for the phase.

The stacked and smoothed residual amplitudes and phases are shown in
Figure~\ref{fig:aver_ampls+phases} (panels (a) and (c)). A zoomed-in view
of the time-series, normalized to the rms scatter, is presented on the
right of this figure (panels (b) and (d)). In panel (b), the zero level
corresponds to the average normalized amplitude of the unobscured
signal. Also shown in panels (b) and (c) are the amplitudes and phases
with the time axis reversed. This is to highlight asymmetries between
the immersion and emergence sides.  The intensity minima caused by the
obscuration of the background source, and the Arago--Poisson spot
between these minima are detected at levels exceeding $10\,\sigma$ in
the amplitude curve. The maximum phase shift associated with the
deepest occultation is 1.3 radians ($\sim 30\,\sigma$). The amplitude
maxima on both sides of the deep minima, corresponding to the first
intensity maxima of the diffraction pattern, are detected at levels of
3~$\sigma$ (immersion) and 5~$\sigma$ (emergence). The depths of the
phase minima occurring at the same time, are approximately 8~$\sigma$
(immersion) and 6~$\sigma$ (emergence). The shapes of the amplitude
bumps and the phase dips are different on the immersion and emergence
sides, the difference being more prominent in the phase curve. The
difference is also visible in the amplitude versus phase plot shown in
panel e of Figure~\ref{fig:aver_ampls+phases}. In this diagram, the
center of the occultation curve, with the maximum phase shift lies on
the right, and the unobscured baselines before and after the
occultation correspond to the concentration of points around phase 0,
amplitude 1. The visibilities on the immersion and emergence sides of
the occultation are plotted with blue and red symbols, respectively.

\begin{figure*}[htbp]
\figurenum{4}
\unitlength=1mm
\begin{picture}(160,90)
\put(0,0){
\begin{picture}(0,0) 
\includegraphics[width=16.0cm]{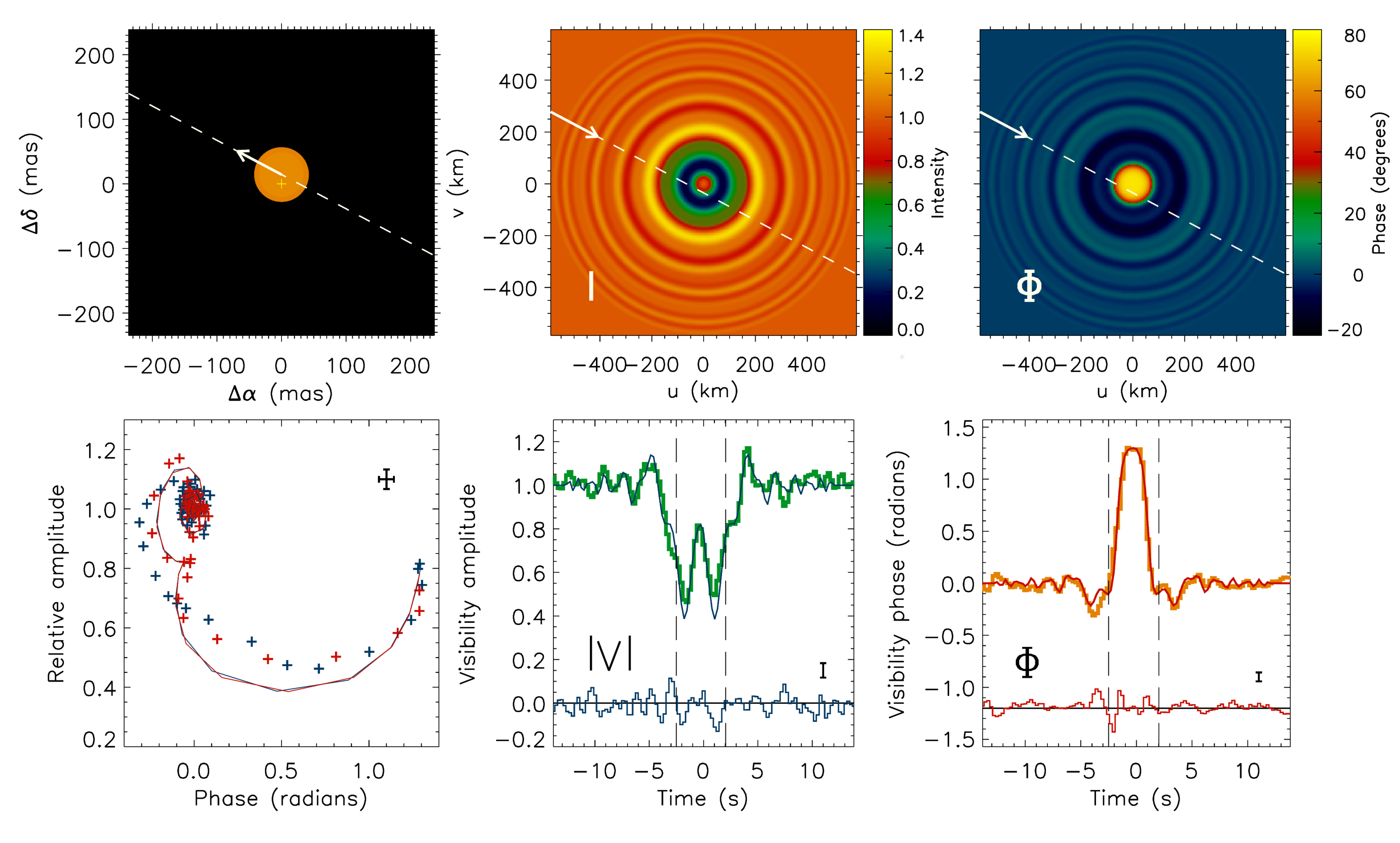}
\end{picture}}
\put(21,90.5){\makebox(0,0){\large \bf \color{white} a)}}
\put(69,90.5){\makebox(0,0){\large \bf \color{white} b)}}
\put(120,90.5){\makebox(0,0){\large \bf \color{white} c)}}
\put(19.5,46.0){\makebox(0,0){\large \bf \color{black} d)}}
\put(69,46.0){\makebox(0,0){\large \bf \color{black} e)}}
\put(120,46.0){\makebox(0,0){\large \bf \color{black} f)}}
\end{picture}
\caption{Top row: opaque circular disk (a) and the intensity (b) 
and phase (c) patterns of a plane wave diffracted by
  this obstacle.  The radius of the occluder and its shortest
  projected distance from the background source (plus sign) are
  adjusted to agree with the occultation curves produced by Palma.
  The silhouette is drawn on the celestial sphere, whereas the
  diffraction patterns are projections onto a plane perpendicular to
  the direction of the radiation source.  The dashed line in panel
  (a) shows the path of the asteroid in the sky relative to the
  background source. The dashed lines in panels (b) and (c)
  indicate the cross section of the diffraction pattern measured at
  Brewster. Bottom row: visibility amplitude versus phase (d) and
  the amplitude (e) and phase (f) profiles along the
  dashed lines shown in the top row. In panel (d), the observing
  points corresponding to the immersion and the emergence are
  indicated with blue and red, respectively. In the calculated
  diagram, these two sides overlap perfectly, because the obstacle is
  symmetric. The observed time-series of the visibility amplitudes and
  phases during the occultation of Palma are shown with thick green
  and orange lines. The time range ($\sim 30$~s) corresponds to the
  spatial range ($\sim 1200$\,km) of the images. The residuals after
  subtracting the model from the observations are shown in the bottom
  of panels (e) and (f). The residuals in phase are
    shifted down by 1.2 radians for clarity (panel (f)). The rms
    scatters of the normalized amplitude and the phase are indicated
    with vertical bars in panels (e) and (f), and with a cross in panel
    (d).}
\label{fig:disk} 
\end{figure*}
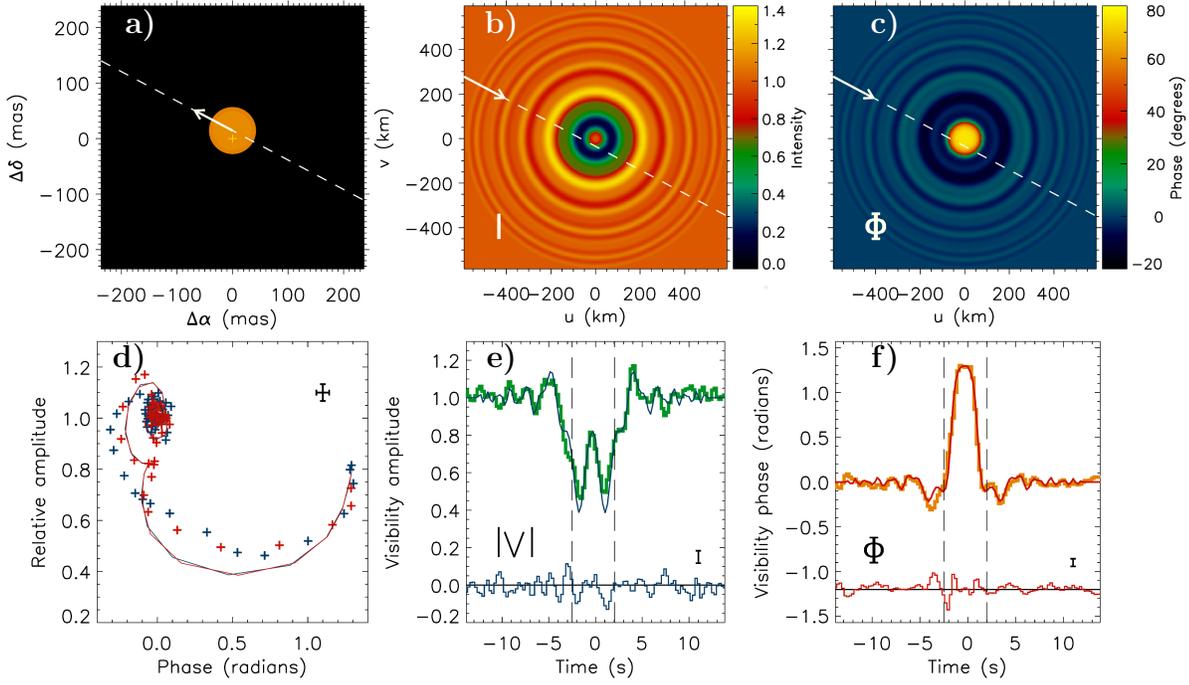

\section{Predicted Visibilities from Silhouette Modeling}

\subsection{Circular Disk}

We first model the occultation curves assuming that the occluder is a
circular disk. In the case in which the background source is
point-like, implying that the incident wavefront is planar, the
formulae for the complex diffracted amplitude are given, for example,
by \cite{1987AJ.....93.1549R} (Eqs. B2 and B4 in their Appendix B)
and by \cite{2013EAS....59...37A} (their Sect. 5.3). 
The finite size of the background source can give rise to a smoothing 
effect \citep{2008ssbn.book..545R}, but in the present
observations this effect is negligible; 99\% of the flux
comes within an angular radius of $\sim 3$ mas, which is much smaller
than the angular Fresnel scale, $\sqrt{\lambda/2d} \sim 40$ mas.
 For this estimate we reprocessed the VCS-II/BG219D VLBA segment observed on
2014 June 9 \citep{2016AJ....151..154G}, and produced the 8.7~GHz
image of 0141+268 shown in Figure~\ref{fig:0141+268}.  To examine the
magnitude of this smoothing effect, we calculated the amplitude and
phase curves from a circular occluder using the actual radio map, and
found no visible difference from the calculation that used a point
source. 

The three input parameters of the circular disk model are the asteroid
radius, $a$, the perpendicular distance, $\Delta$, of the
  Brewster antenna from the center of the shadow path, and the time,
$t_0$, of the closest approach. The set of input parameters that could produce 
the closest match to the observed visibility amplitudes and phases
was determined by finding the minimum of a reduced $\chi^2$ function.
We compare both amplitudes and phases to the model predictions, and
define the reduced $\chi^2$ function by
\begin{equation}
\begin{array}{lcl}
\chi^2 &=& \frac{1}{n-p}\sum_{i=1}^n \left( \frac{|V|_i^{\rm obs} - |V|_i^{\rm calc}}{\sigma_i^{|V|}}\right)^2 \\ 
    & &  + \frac{1}{n-p}\sum_{i=1}^n \left( \frac{\Phi_i^{\rm obs} - \Phi_i^{\rm calc}}{\sigma_i^{\Phi}}\right)^2 \; ,
\end{array}
\label{eq:chi2}
\end{equation}
where $n$ is the number of the data points, $p$ is the number of free
parameters (here 3), $|V|_i$ is the visibility amplitude, and $\Phi_i$
is the visibility phase. The superscripts ``obs'' and ``calc'' refer
to the observed and calculated values, respectively. For
  acceptable models, the reduced chi-square, that is, chi-square
  divided by the degrees of freedom as defined in Eq.~\ref{eq:chi2}
  should be less than 2.5 (see, e.g., \citealt{1976ApJ...208..177L}).

The values of the radius, and the distance and time of the closest
approach found by this minimization are $a=95.9\pm 2.4$\,km,
$\Delta=-31.6\pm 3.4$\,km, and $t_0 =$ UTC 14:31:19.63 $\pm 0.07$~s
(Coordinated Universal Time). The solution obtained, with the
  reduced chi-square $\chi^2= 3.76$, is not particularly good. The
  parameters of this and other models tested in the present paper are
  listed in Table~\ref{tab:shapes}.

\setcounter{table}{0}
\begin{table*}[t!]
\renewcommand{\thetable}{\arabic{table}}
\centering
\caption{Best-fit parameters of silhouette models} \label{tab:shapes}
\begin{tabular}{llllll}
\tablewidth{0pt}
\hline
\hline
Model & $a_{\rm eff}$ & $\Delta$ & $t_0$  & Other Parameters & $\chi^2$ \\
      & (km)        & (km)     &  UTC 14:31 + &   &  \\ \hline
circle  & $95.9\pm2.4$ & $31.6\pm3.4$ & $19.63\pm 0.07$~s & & 3.76 \\
polyhedral model 1 & $95.8\pm2.7$ & $26.1\pm 3.9$ & $19.66\pm0.07$~s & & 4.61 \\
polyhedral model 2 & $97.2\pm2.1$ & $34.7\pm2.7$ & $19.57\pm0.07$~s & & 6.01 \\
model 1 rotated by $180\degr$ & $96.3\pm2.7$ & $25.1\pm 3.9$ & $19.69\pm0.07$~s & & 3.53 \\
ellipse & $96.1\pm2.4$ & $25.4\pm4.5$ & $19.68\pm0.07$~s & $e=0.40 \pm
0.13$, P.A.$=-3\fdg1 \pm {16\fdg3}^a$ & 3.59 \\ 
average random model & $96.0^b$ & $19.7\pm4.8$ & $19.76\pm0.08$~s & $r=0.88$ & 2.89 \\ \hline
\end{tabular}

\noindent
{\footnotesize $^a$ position angle with respect to the declination axis; $^b$ The effective radius was kept constant in the Markov chain Monte Carlo simulation.}  
\end{table*}

The predicted diffracted intensity and phase patterns for a circular
obstacle with the quoted radius are shown in Figure~\ref{fig:disk}
(panels (b) and (c)).  The predicted visibility amplitudes and phases at
the best-fit perpendicular distance from the shadow centre are shown
below them (panels (e) and (f)), together with the observed amplitudes and
phases. The corresponding amplitude vs. phase diagrams are shown in
panel (d) of this figure.

In addition to the circular shape and the two models from light-curve
inversion, we tested elongated and asymmetric models, constructed
using circles or ellipses. The following silhouette shapes were
tested: ellipse, merged binary, and crescent. An ellipse with
an eccentricity $e=0.4$ and an effective radius of $a_{\rm eff}=96$\,km
can reproduce the observations equally well as the circular model 
(the best-fit parameters are listed in Table~\ref{tab:shapes}), but
the other two models result in worse fits than those obtained using a
circle.


\begin{figure*}[htbp]
\figurenum{5}
\unitlength=1mm
\begin{picture}(160,93)
\put(0,0){
\begin{picture}(0,0) 
\includegraphics[width=16.0cm]{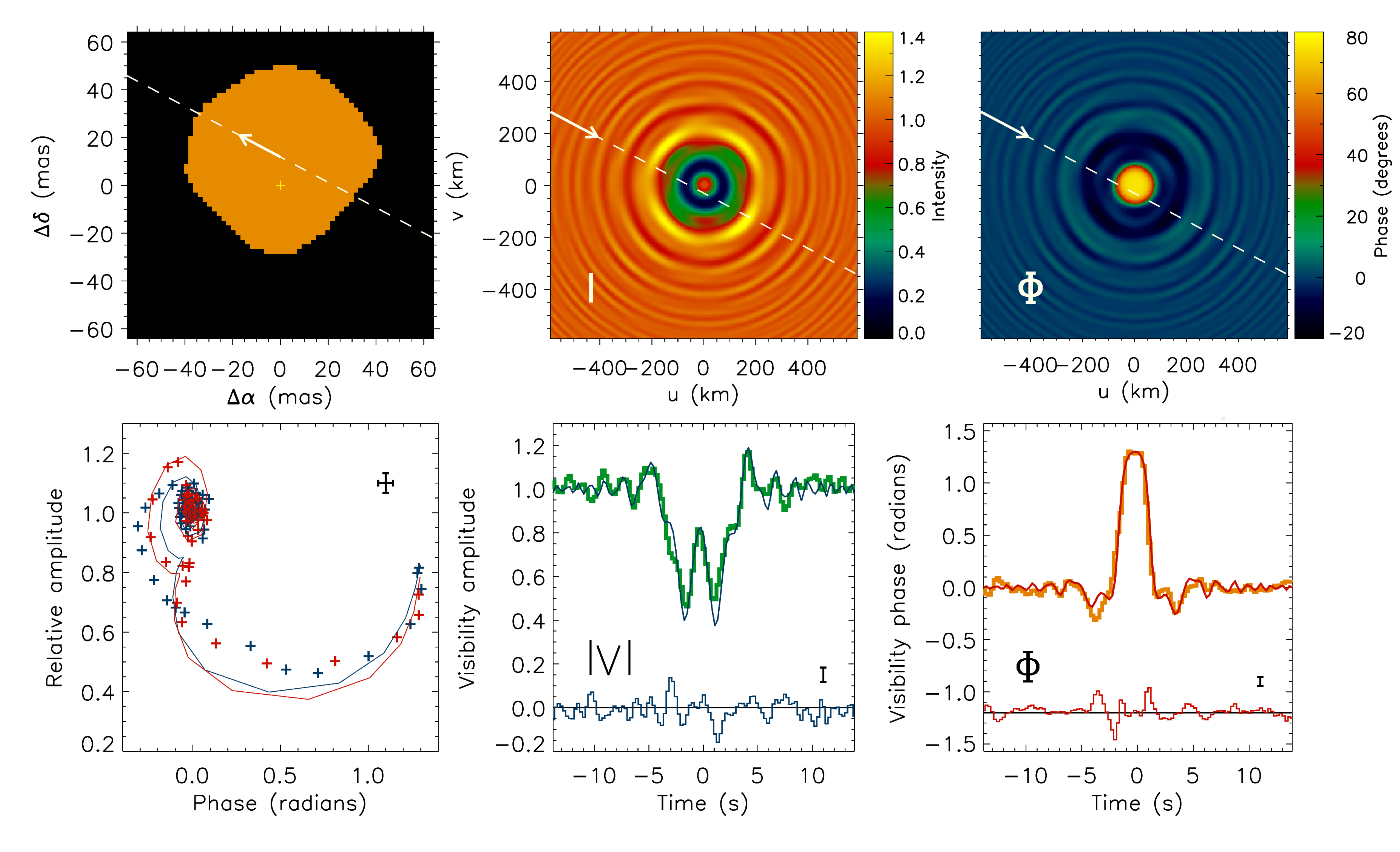}
\end{picture}}
\put(21,90.5){\makebox(0,0){\large \bf \color{white} a)}}
\put(69,90.5){\makebox(0,0){\large \bf \color{white} b)}}
\put(120,90.5){\makebox(0,0){\large \bf \color{white} c)}}
\put(19.5,46){\makebox(0,0){\large \bf \color{black} d)}}
\put(69,46){\makebox(0,0){\large \bf \color{black} e)}}
\put(120,46){\makebox(0,0){\large \bf \color{black} f)}}
\end{picture}
\caption{Same as Figure~\ref{fig:disk} but the silhouette model is
adopted from the Interactive Service for Asteroid Models (ISAM; see the text). 
Furthermore, a close up of the asteroid silhouette is shown in panel (a), 
so the physical scale is no longer the same as that in panels (b) and (c).}
\label{fig:palma_model}
\end{figure*}
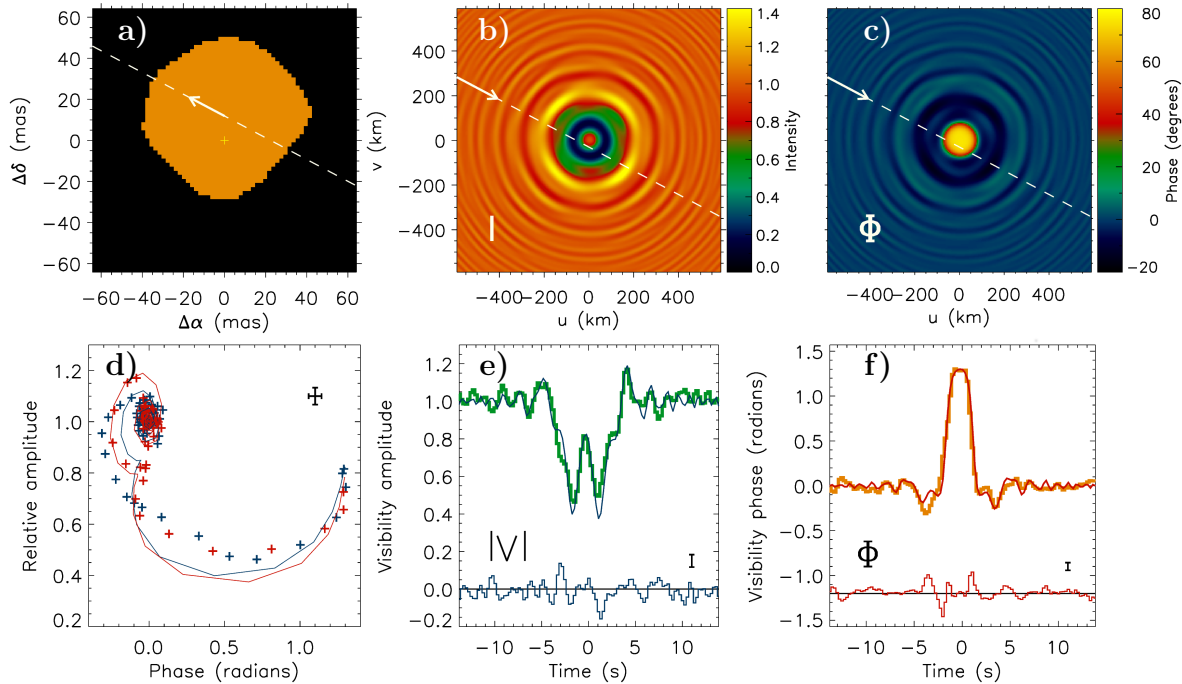

\subsection{Polyhedral Models from Light Curves}

\label{sec:isam_models}

Two three-dimensional models for Palma, with different spin axis ecliptic
coordinates ($\lambda_{\rm P},\beta_{\rm P}$) are available via an interactive
service, ISAM, provided by the Astronomical Observatory, Pozna{\'n}
\citep{2012A&A...545A.131M}\footnote{\url{http://isam.astro.amu.edu.pl}}. The
models are derived from photometric observations using light-curve
inversion (\citealt{2001Icar..153...37K};
\citealt{2011Icar..214..652D}; \citealt{2011A&A...530A.134H}). The
predicted silhouettes of Palma at the time of the occultation were
used to calculate intensity and phase maps. The silhouette prediction
takes into account the light travel time (approximately 28.5
minutes). The rotation period of Palma is approximately 8.6\,hr.
 
The diffraction patterns were calculated using a method adopted from
\cite{2014ApOpt..53.3540T}. In this method, the silhouette is
presented as a grid of rectangles, and the complex amplitude of the
diffracted wavefront is obtained as the sum of those caused by
individual rectangles.  The resulting complex amplitude is given in
Eq. (10) of \cite{2014ApOpt..53.3540T} (inside the modulus bars).  

The polyhedral models are given in the celestial coordinate system.
These models are asymmetric, and the angle at which the asteroid
crossed the radio source affects the occultation curve.  The apparent
motion of the asteroid in the sky relative to the radio source was
directed from the southwest to the northeast; the tilt angle with
respect to the hour circle passing through the radio source was
approximately $62\degr$ ($152\degr$ measured counterclockwise from the
declination circle). The orientation of the measured cross section of
the diffraction pattern was also indicated for the circular disk model
(Figure~\ref{fig:disk}), although for a symmetric occluder the
obliqueness is of no consequence.  The tilt angle is determined by the
orientation of the ecliptic and the inclination of the asteroid orbit
with respect to this. Because the inclination is assumed to be known
to a high accuracy, the tilt angle is kept constant in the present
calculations.

We fitted the observed occultation curves with the shapes from
light-curve inversion by varying the effective radius of the model, and
the distance and the time of the closest approach. The effective
radius is defined by $a_{\rm eff}=\sqrt{A/\pi}$, where $A$ is the
projected area of the asteroid. The diffraction patterns and the
predicted occultation curves for one of the asteroid models (with the
spin vector ecliptic coordinates $\lambda_{\rm P}=221\degr, \beta_{\rm
  P}=-47\degr$; we call this model 1)
are shown in Figure~\ref{fig:palma_model}.  The best-fit parameters of
this model and the other polyhedral model (``model 2''), with
$\lambda_{\rm P}=44\degr, \beta_{\rm P}=17\degr$, are listed in
Table~\ref{tab:shapes}.

Compared with the circular disk model, the polyhedral models do not
improve the overall agreement between the observed and predicted
amplitude and phase curves. Model 1 reproduces reasonably well the
amplitude maximum on the emergence side (corresponding to the arch of
red plus signs in the top left of the amplitude vs. phase diagram),
but neither of the models can reproduce the prominent dent in the
phase on the immersion side (the bight of blue crosses on the 
left). Both the circular disk and the adopted polyhedral models
produce overly deep amplitude minima on both sides of the
Arago--Poisson spot. In the amplitude vs. phase diagram, the effect is
that the predicted curves lie below the observed points in the middle
of the plot (between phases 0 and 1 radian). 

The polyhedral model 1, when turned by $180\degr$ about the spin
  axis, corresponding to the asteroid silhouette half a rotation
  period earlier or later than the model shown in
  Fig.~\ref{fig:palma_model}, gives a better agreement than the original model. 
  This is discussed briefly in Sect.~\ref{sec:random_average}. The best-fit
  parameters of the rotated model are given in Table~\ref{tab:shapes}.

\begin{figure}[htbp]
  \figurenum{6} 
\unitlength=1mm
\begin{picture}(80,65)
\put(0,0){
\begin{picture}(0,0) 
\includegraphics[width=8.0cm]{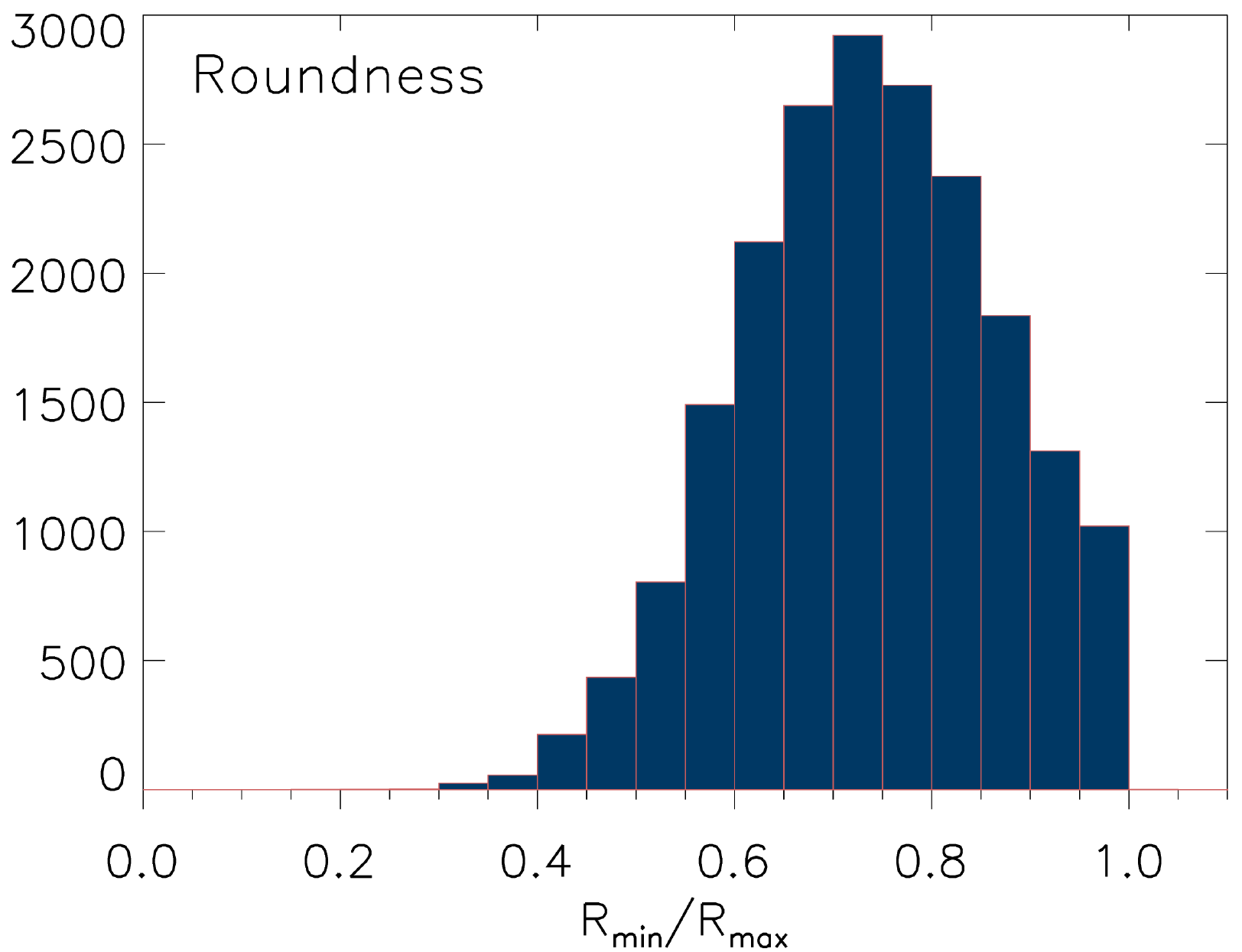}
\end{picture}}
\end{picture}
\caption{Distribution of the ratio of the inscribed and circumscribed
  radii for a sample of 20,000 contiguously random shapes, constructed
  by a Markov chain Monte Carlo method described in the text. The
  histogram is assumed to correspond to the probability distribution
  of the roundness parameter of the models explaining the amplitude
  and phase curves observed during the radio occultation by Palma.}
\label{fig:roundness}
\end{figure}

\vspace{2mm}

\subsection{Estimate of the Roundness from Random Models}
\label{sec:random_shapes}

Shapes obtained from light-curve inversion and the best-fit elliptical
model suggest that the asteroid silhouette deviates from a perfect
circle. When the roundness of the shape is characterized by the ratio,
$r$, of the inscribed and circumscribed circles, the roundness of the
best-fit polyhedral model is $r=0.83$, whereas for the best-fit
elliptical model ($e=0.4$) this ratio is $r=0.92$.  We estimate the
roundness by generating a sequence of random contiguous silhouettes
using the Metropolis--Hastings algorithm (\citealt{1953JChPh..21.1087M}; 
 \citealt{Hastings1970}). The distribution of the
roundness parameter of random silhouettes obtained in this manner
should approximate the probability distribution of the roundness of
Palma.

In the silhouette models we only varied the ratio of the minimum and
maximum circles that just fit inside and enclose the shape, but kept
the effective radius, that is, the surface area of the asteroid,
constant. We used the best-fit value, $a_{\rm eff}=96$\,km, from the
simulations described above. In each step of the sequence, the
roundness parameter $r^\prime$ was picked at random from a Gaussian
distribution with a width of 0.235 ($\sigma=0.1$), centered on the
previously accepted value $r$. The area inside the minimum radius was then
completely filled, and the area between the minimum and maximum radii
was filled with randomly placed pixels, however so that the result was
contiguous.

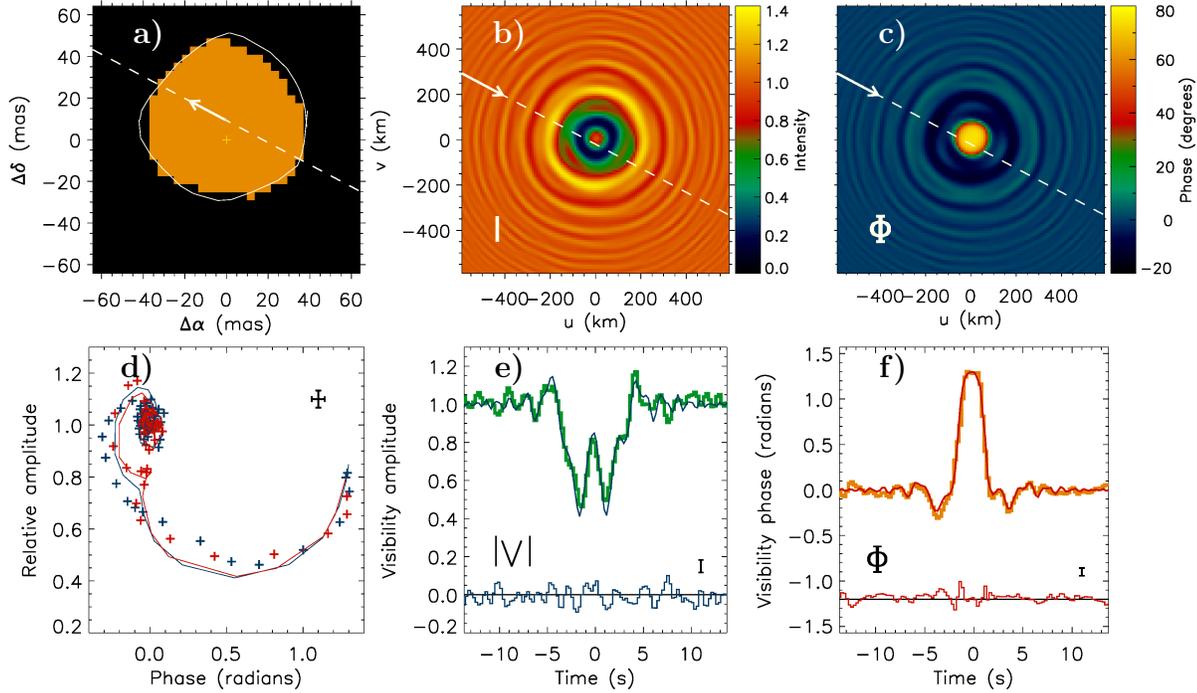
\begin{figure*}[htbp]
\figurenum{7}
\unitlength=1mm
\begin{picture}(160,90)
\put(0,0){
\begin{picture}(0,0) 
\includegraphics[width=16.0cm]{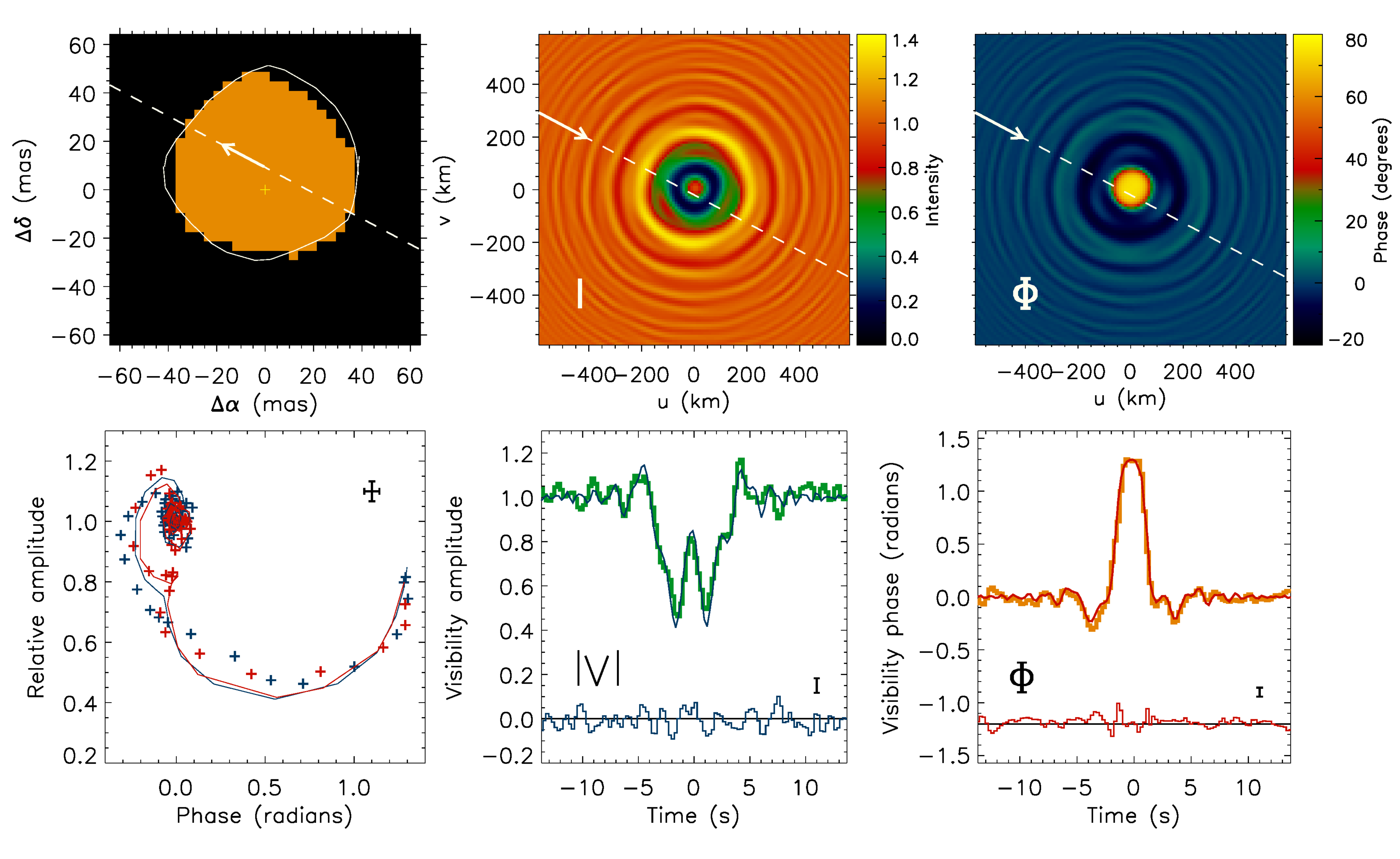}
\end{picture}}
\put(21,90.5){\makebox(0,0){\large \bf \color{white} a)}}
\put(69,90.5){\makebox(0,0){\large \bf \color{white} b)}}
\put(120,90.5){\makebox(0,0){\large \bf \color{white} c)}}
\put(19.5,46){\makebox(0,0){\large \bf \color{black} d)}}
\put(69,46){\makebox(0,0){\large \bf \color{black} e)}}
\put(120,46){\makebox(0,0){\large \bf \color{black} f)}}
\end{picture}
\caption{Same as Figure~\ref{fig:palma_model} but the silhouette
  model is the average of the best-fitting random models (see the 
  text). The silhouette from the ISAM service, half a rotation period
  before (or after) the time of the occultation, is superposed on the
  averaged random model (panel (a), see the text).}
\label{fig:random_average}
\end{figure*}

The $\chi^2$ value of the best-fit cross section of the diffraction
pattern was used as a measure of the probability of the model, $\ln P(r)
\propto -0.5\,\chi^2$, and in accordance with the Metropolis--Hastings
principle, the acceptance ratio, $\ln P(r^\prime) - \ln P(r)$, was
compared with the logarithm of a uniform random number on the interval
$[0,1]$ when deciding whether to accept or reject the candidate
$r^\prime$.  The roundness distribution of 20,000 realizations of random
shapes constructed in the course of this procedure is shown in
Figure~\ref{fig:roundness}. The distribution is skewed to the left
because $r$ cannot exceed unity.  The moments of the distribution are
${\mu_r}=0.74$ (mean), $\sigma_r =0.13$ (standard deviation), and
$\gamma_1 = -0.16$ (skewness). The distribution suggests that the
asteroid shape deviates from a perfect circle by $26\pm 13\%$. The
roundness parameter is greater than 0.65 with 95\% confidence.

\subsection{Best Random Models}
 
\label{sec:random_average}

Many of the random models constructed in the course of the Markov
  chain reproduce the observations better than any of the geometrical
  models or the polyhedral models tested above.  One of the random
  models giving a good agreement with the occultation curves ($\chi^2
  = 2.24$) is shown in Figure~\ref{fig:random_palma} of Appendix~A.
  As discussed in Section~\ref{sec:discussion}, with the pixel size
  10~km used in this simulation, one cannot obtain a unique solution
  for the asteroid shape. One can notice, however, that the best
  solutions resemble each other. In Figure~\ref{fig:random_average} we
  present a fit to occultation curves using the average of random
  models with $\chi^2 \le 2.5$ (altogether 264). In the average
  silhouette, pixels near the boundaries can have values between 0 and
  1. In this model we have assumed pixels with values $> 0.5$ are
  fully opaque.  Because the effective radius in the simulation of
  random shapes was fixed, the only parameters fitted here are the
  distance and the timing of the closest approach. The averaging has a
  rounding effect; the roundness parameter for this model is
  $r=0.88$. The best solution for the timing and the perpendicular
  distance from the shadow center gives $\chi^2=2.89$.  

  The average silhouette resembles the polyhedral model 1 with
  $\lambda_{\rm P}=221\degr, \beta_{\rm P}=-47\degr$ from the ISAM
  service, taken half a rotation period ($4^{\rm h}17^{\rm m}27^{\rm
    s}$) before or after the time of the closest approach. This
  corresponds to turning the shape solution by $180\degr$ about the
  spin axis.  To illustrate this coincidence, we have drawn the outlines
  of the ISAM model 1 for the time 10:13:56 UTC in panel (a) of
  Figure~\ref{fig:random_average}.  This silhouette is practically
  identical with the prediction for 18:48:50 UTC. The best-fit
  parameters of the rotated model are given in
  Table~\ref{tab:shapes}. The difference of half a period between the
  best-fit time and the time of the closest approach suggests that the
  polyhedral model 1 is a good approximation of the true shape, but
  the initial rotation angle of the solution needs to be corrected.

\section{Discussion}
\label{sec:discussion}

\begin{figure}[htbp]
  \figurenum{8} 
\unitlength=1mm
\begin{picture}(80,105)
\put(0,0){
\begin{picture}(0,0) 
\includegraphics[width=8.0cm]{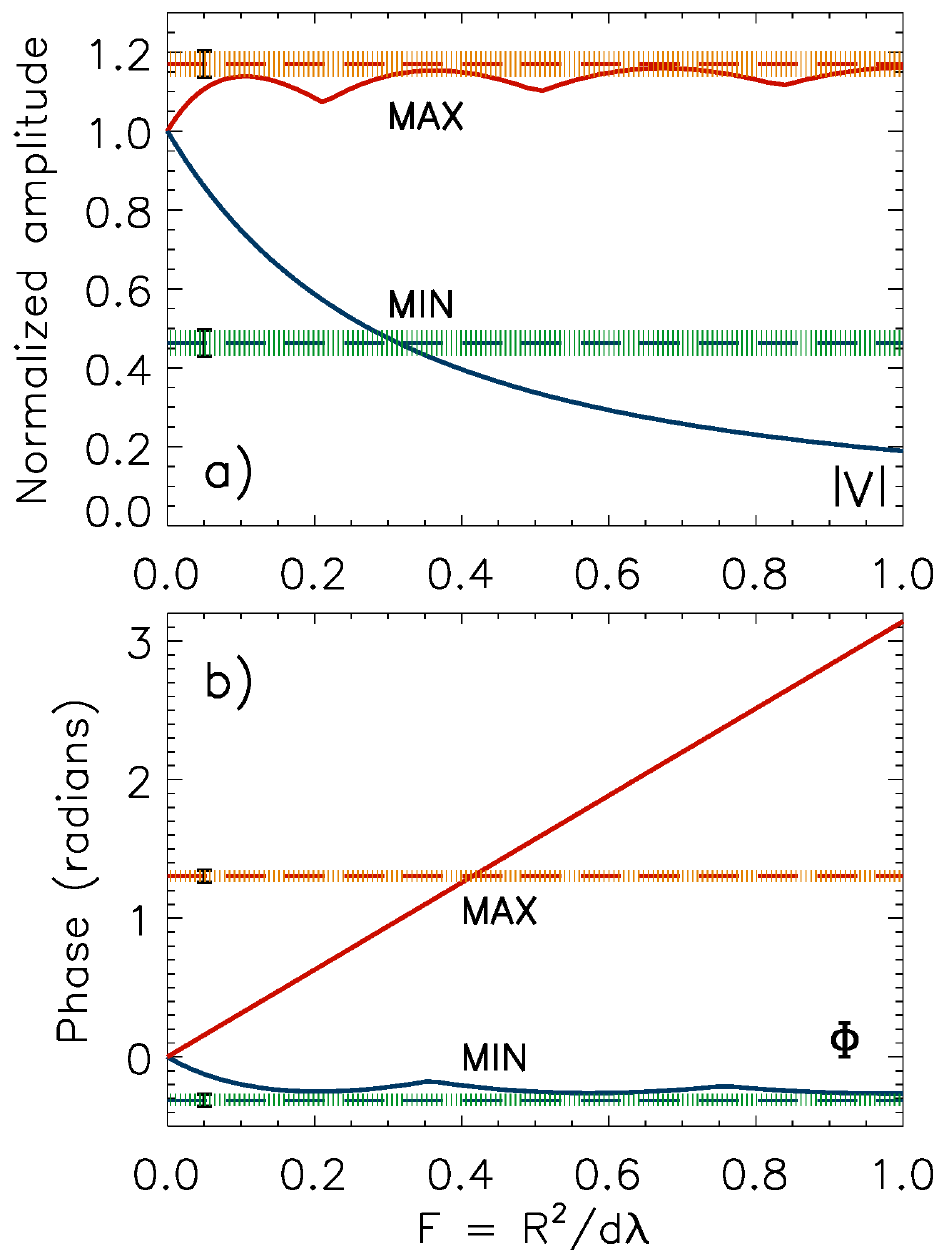}
\end{picture}}
\end{picture}
\caption{Maximum and minimum amplitude (a) and phase shift (b) of the
  diffracted wavefront as functions of the Fresnel number, $F=R^2/(d
  \lambda)$, for a circular disk with the radius $R$. The other
  parameters here are the distance, $d$, and the wavelength,
  $\lambda$. The hatched horizontal bars describe the observed 
maximum and minimum values and their 1$\sigma$  errors.}
\label{fig:famph}
\end{figure}

\subsection{Size and Shape of the Asteroid}

The possibility of measuring both the amplitude and the phase of the
diffracted wavefront increases significantly the accuracy of asteroid
sizing as compared with single-dish observations. The maximum phase
shift is particularly sensitive to the effective diameter of the
asteroid. This can be seen in Figure~\ref{fig:famph}, which shows the
maximum and minimum amplitudes and phases of wavefronts diffracted by a
circular obstacle as functions of the Fresnel number, $F$. The phase
shift is directly proportional to $F$, with the proportionality
constant $\pi$. The minimum and maximum curves in
Figure~\ref{fig:famph} represent measurements through the shadow's
center, whereas the observed cross section of the diffraction pattern
is off from the center. Because the phase pattern is flat topped (see
Figure~\ref{fig:disk}), a small offset does not affect the maximum
phase. In contrast, the Arago--Poisson spot is strongly peaked, and the
strength of the central maximum decreases rapidly with the distance
from the shadow center. All silhouette models considered here give
practically the same best-fit effective diameter for Palma.  This
diameter, $192.1\pm2.3$\,km, corresponding to the maximum phase shift
$76\fdg4\pm2\fdg5$, agrees very well with recent determinations
from thermal emission at infrared wavelengths
(\citealt{2012P&SS...73...98C}: $191.1\pm 2.7$\,km;
\citealt{2014ApJ...791..121M}: $186.5 \pm 6.3$\,km).

Even though the exact shape of the occluder is difficult to discern in
Fraunhofer diffraction, deviations from circular shape can leave
recognizable traces to the occultation curves. The amplitude
vs. phase plot is useful for identifying asymmetries. Characteristic
of the diagram derived from the present observations is a kink seen
only on the emergence side, and a prominent depression of phases on
the immersion side. These features rule out the circular model, 
and any models where the leading and trailing edges mirror each other. 

Several random shapes, generated in the course of the Markov chain
Monte Carlo sequence (Section~\ref{sec:random_shapes}) and having
  the same projected area as the best-fit circular and elliptical
  models, give acceptable fits to the occultation curves. One example
is shown in Figure~\ref{fig:random_palma}.  These random shapes are
too irregular to offer a plausible model for a 200\,km diameter
asteroid with substantial surface gravity. The average of the
  best random shapes shown in Figure~\ref{fig:random_average} is,
  however, realistic. The average of random models produces larger
  residuals than the best individual random shapes, but the residuals
  are clearly smaller than those for the other models tested here. The
  average model therefore gives a plausible approximation for the
  shape of the silhouette at the time of the occultation.  It
  resembles one of the polyhedral models from light-curve inversion,
  rotated by $180\degr$ about the spin axis. This suggests that the
  initial rotation angle (or the initial epoch) of the shape solution
  from light curves needs to be adjusted.

  Also, optical occultation observations show some inconsistency 
   with the models from light-curve inversion. Several observations 
  at visual wavelengths are archived at NASA's Planetary Data
  System\footnote{\url{https://sbn.psi.edu/pds/resource/occ.html}}
  \citep{dunham2017}. One of the events, observed on 2007 Jan 26 UT
  9:50, with 20 chords across the asteroid, gives a good idea of the
  asteroid
  shape\footnote{\url{https://sbnarchive.psi.edu/pds3/non_mission/EAR_A_3_RDR_OCCULTATIONS_V14_0/document/372palma2007jan26.png}}.
  The compatibility of the three-dimensional models 1 and 2 with these
  observations can be assessed by inspection of plots made available
  on the website of the Database of Asteroid Models from Inversion
  Techniques (\citealt{2010A&A...513A..46D}). These plots
  indicate that model 1 agrees better with the observations than model
  2, but even for model 1 the predicted silhouette does not correspond
  exactly to the observed shape. The occultation chords suggest that the
  northwestern side of the silhouette was rather square at the time 
of this observation, whereas the predicted shape is tapered on this 
side\footnote{\url{http://astro.troja.mff.cuni.cz/projects/asteroids3D/data/archive/1-1000/A224.M298.occ_2007-01-26.pdf}}.

\cite{2014ApOpt..53.3540T} suggested a technique of pixel-by-pixel
  reconstruction, which, together with constraints on the smoothness
  of the surface, can be used for determining the shape of an asteroid
  from occultation observations with low Fresnel numbers. A single cut
  through the diffraction pattern, as observed here, is probably
  insufficient to provide a unique solution.  According to
  \cite{2014ApOpt..53.3540T}, it is possible to reconstruct the shape
  of an occluder with several measurements of the intensity profile
  through the shadow pattern.  They estimate that an unambiguous
  recovery of the silhouette requires twice as many measurement points
  (number of apertures times number of samples per aperture) than
  there are pixels in the silhouette image. The silhouette models used
  here have approximately 300 pixels (with a pixel size of
    10\,km), while the total number of useful measurement points (in
  amplitude and phase) is $\sim 160$. Decreasing the number of
    pixels in the silhouette model to correspond to the number of
    observation points (requiring here a pixel size of $\sim 20$\,km)
    results in a model too coarse to be helpful.

  Therefore, radio interferometric
    measurements also should cover different parts of the diffraction pattern to
  provide for a unique solution for the asteroid shape.  The spacing
  between telescopes across the shadow path should be of the order of
  the spatial Fresnel scale, $\sqrt{d\lambda/2}$, which in the case of
  the present occultation by Palma, is $\sim 100$~km.  Currently,
  such spacings would be available with VLBI networks, such as the
  VLBA and its counterparts in other continents, in the fortuitous
  event that the shadow of an asteroid should pass over two or more of
  their antennas.

\subsection{Implications for the Asteroid Orbit}

The measurement of the occultation curves in the Fraunhofer regime
gives an estimate for the position of the asteroid at the time of the
deepest occultation. Particularly, when the Arago--Poisson spot is detected,
the time of closest approach, and the positional offset between the
radio source and the center of the asteroid at that time can be
determined rather accurately.

\begin{figure}[bthp]
  \figurenum{9}
  \includegraphics[width=\columnwidth]{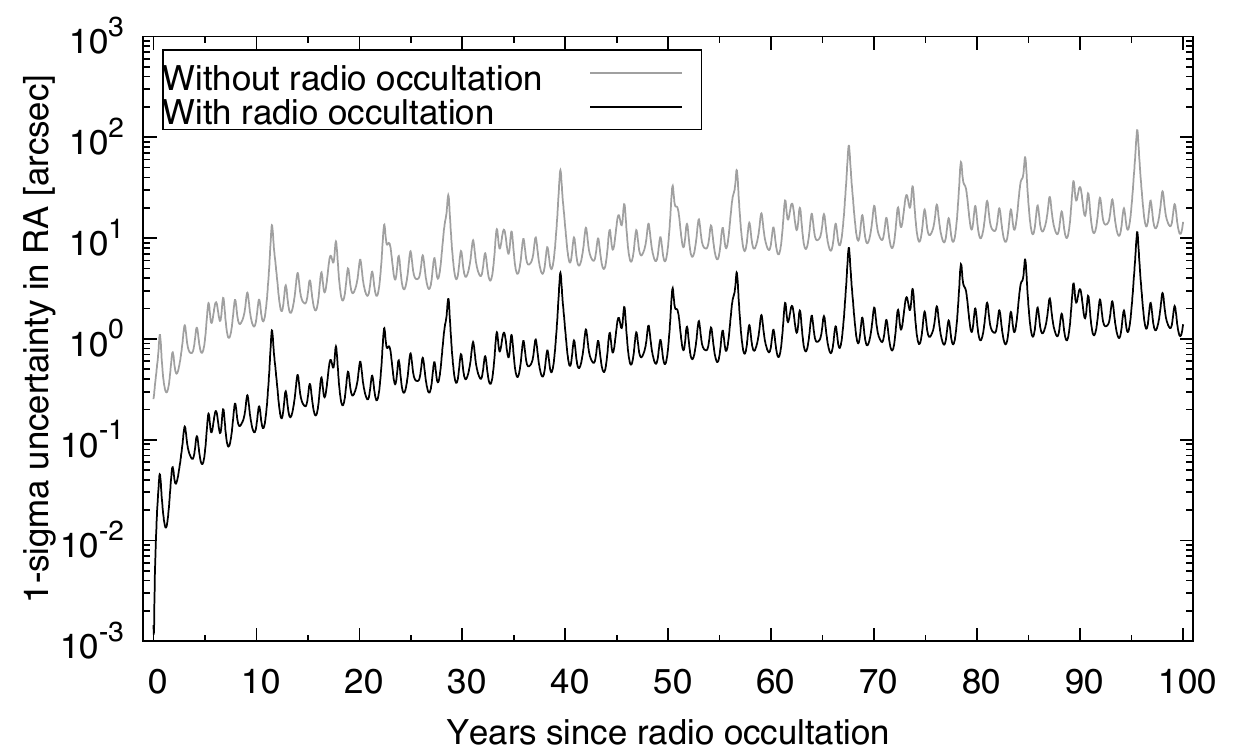}
  \includegraphics[width=\columnwidth]{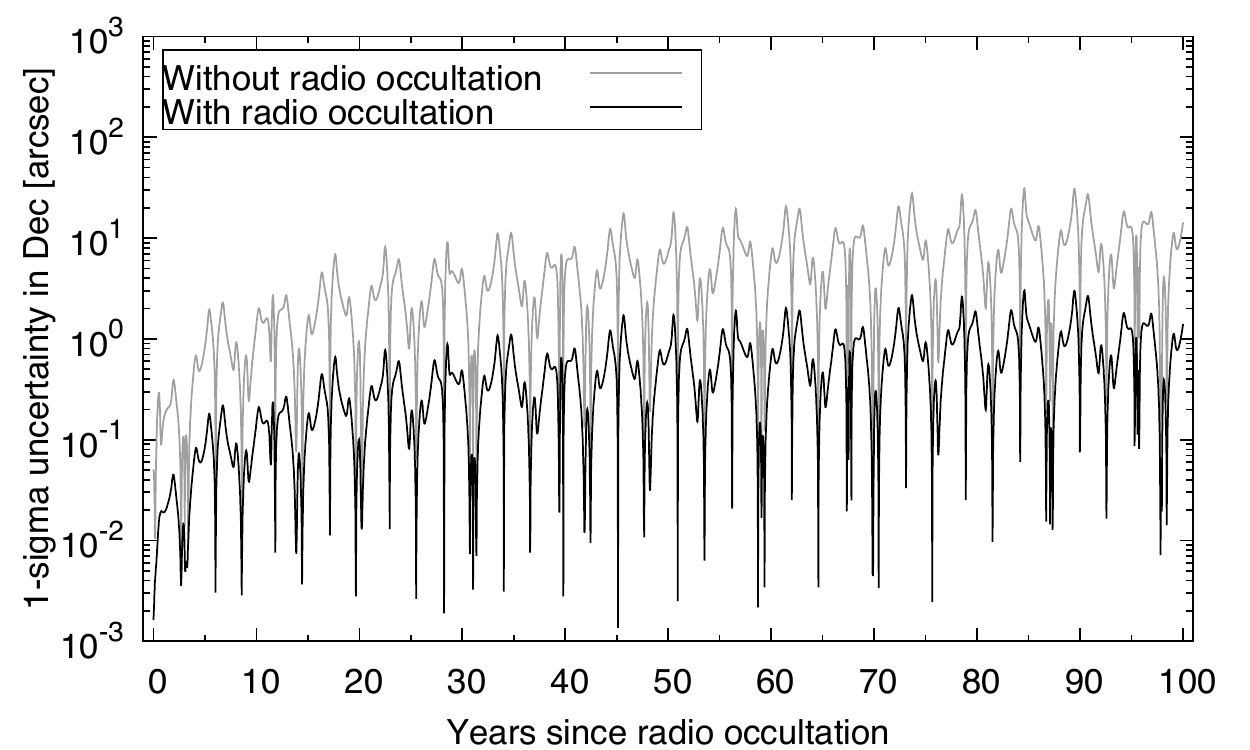}
  \caption{Ephemeris uncertainty for (372) Palma for 100 years
    subsequent to the occultation in (top) R.A. and (bottom) decl. for a
    hypothetical geocentric observer. The black line refers to the
    prediction that includes the occultation-based astrometry, whereas
    the gray line is a reference that does not include information
    about the occultation. The long-term trend in the ephemeris
    uncertainty is caused by the increasing orbital uncertainty and
    the short-term fluctuation is caused by the varying distance from
    the Earth to the asteroid (the peaks correspond to the shortest
    distances).}
\label{fig:ephunc}
\end{figure}

The measured apparent coordinates for Palma on 2017 May 15 
  14:31:19.76 UTC as seen from {\sc br-vlba} are R.A. 
  01:44:33.5537 and decl. 27:05:03.125.  Both coordinates have an
uncertainty of $0\farcs002$, mostly due to uncertainties in the
topocentric distance to Palma and the timing of the closest encounter
($\Delta t_0 = 0.08$~s). The position of 0141+268 is known with an
order of magnitude higher accuracy; according to the RFC the uncertainties 
of the R.A. and decl. coordinates are 0.21
mas and 0.28 mas, respectively. The ephemeris by the Jet Propulsion
Laboratory's HORIZONS system for the observing date is in agreement
with the measurement given a 3$\sigma$ ephemeris uncertainty of about
$0\farcs05$. Including the coordinates derived from the radio
occultation should reduce the uncertainty of the orbital solution,
given that the ephemeris uncertainty is an order of magnitude larger
than the measurement uncertainty. To test this hypothesis we computed
an orbital solution for Palma with all existing optical astrometry
available through the Minor Planet
  Center\footnote{\url{https://minorplanetcenter.net/iau/mpc.html}}
  (excluding visible-wavelength occultations) and the radio
  occultation included using the OpenOrb software
  \citep{2009M&PS...44.1853G}. We deliberately excluded
  visible-wavelength occultations to get a better understanding of how
  much improvement can be obtained by adding a single
  radio occultation measurement to regular astrometry spanning more
  than a century.  We used the linearized least-squares method with
outlier rejection and included gravitational perturbations by all
8 planets and the 25 most massive asteroids, as well as first-order
relativistic corrections for effects caused by the Sun. For the
observational error model we used that by Baer et al. without
correlations (\citealt{2011Icar..212..438B};
\citealt{2017AJ....154...76B}). Although the more than 1600
astrometric observations of Palma span more than 120 years (1893 September 29 
to 2017 November 09), the addition of a single radio occultation measurement
has a non-negligible effect on the uncertainties of the orbital
elements (in this particular case, a reduction of up to tens of
  percent) and reduces the uncertainty of ephemerides by an order of
magnitude (Figure~\ref{fig:ephunc}).

\vspace{2mm}

The National Radio Astronomy Observatory is a facility of the National
Science Foundation operated under cooperative agreement by Associated
Universities, Inc. This work made use of the Swinburne University of
Technology software correlator, developed as part of the Australian
Major National Research Facilities Programme and operated under
licence. J.H.and K.M. acknowledge support from the ERC Advanced Grant
No. 320773 ``SAEMPL''. M.G. acknowledges support from the Academy of
Finland (grant \#299543).  We thank the anonymous referee for
  insightful comments that helped to improve this manuscript, Mika
  Juvela for advicing our use of the Monte Carlo simulation, and Dave
  Herald for pointing out the results from optical occultation
  observations.

\bibliographystyle{aasjournal} 

\bibliography{palma}

\appendix

\section{Diffraction Patterns and Occultation Curves from 
a Contiguously Random Shape}

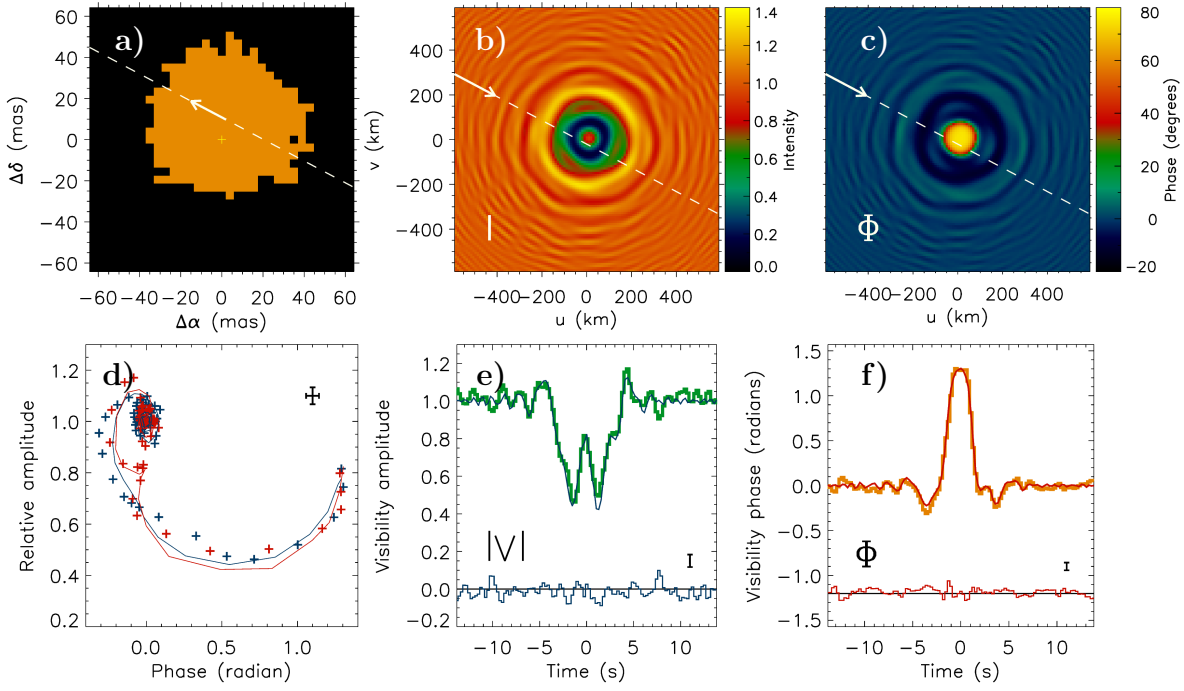
\begin{figure*}[htbp]
\figurenum{10}
\unitlength=1mm
\begin{picture}(160,90)
\put(0,2){
\begin{picture}(0,0) 
\includegraphics[width=16.0cm]{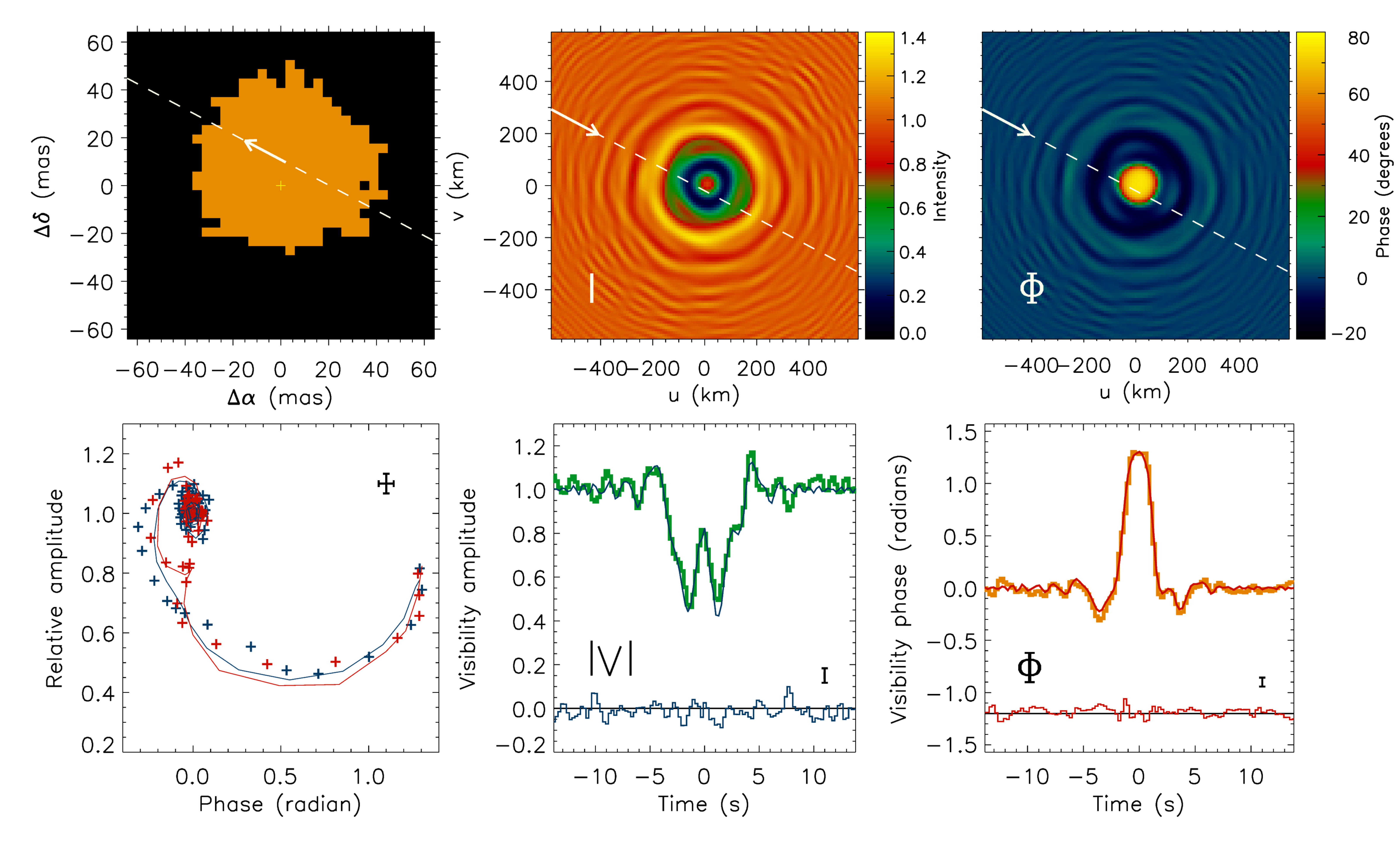}
\end{picture}}
\put(21,90.5){\makebox(0,0){\large \bf \color{white} a)}}
\put(69,90.5){\makebox(0,0){\large \bf \color{white} b)}}
\put(120,90.5){\makebox(0,0){\large \bf \color{white} c)}}
\put(19.5,46){\makebox(0,0){\large \bf \color{black} d)}}
\put(69,46){\makebox(0,0){\large \bf \color{black} e)}}
\put(120,46){\makebox(0,0){\large \bf \color{black} f)}}
\end{picture}
\caption{Random shape producing small residuals with
  $\chi^2=2.24$. The ratio of the inscribed and circumscribed radii
  for this model is $r=0.76$. The panels are the same as those in
  Fig.~\ref{fig:palma_model}.}
\label{fig:random_palma}
\end{figure*}

Figure~\ref{fig:random_palma} shows an example of a random shape that 
reproduces the observations reasonably well.

\end{document}